\documentclass[a4paper,11pt]{article}
\pdfoutput=1
\usepackage{jheppub}
\usepackage[T1]{fontenc}
\usepackage{lmodern,microtype}
\usepackage{amsmath,amssymb,bm,booktabs,array,tabularx,multirow,slashed}
\usepackage{xcolor,graphicx,placeins}

\usepackage[compat=1.1.0]{tikz-feynman}

\hypersetup{colorlinks=true,citecolor=blue,urlcolor=blue,linkcolor=blue,linktocpage=true,pdftitle={Measuring Vcb at electron-proton colliders},pdfauthor={David Marzocca}}
\allowdisplaybreaks

\newcommand{\Vcb}{\lvert V_{cb}\rvert}
\newcommand{\Vub}{\lvert V_{ub}\rvert}
\newcommand{\pt}{p_{\mathrm T}}
\newcommand{\dif}{\mathop{}\!\mathrm{d}}

\title{Measuring $V_{cb}$ at electron-proton colliders}

\author[1]{David Marzocca}

\affiliation[1]{INFN, Sezione di Trieste, SISSA, Via Bonomea 265, 34136 Trieste, Italy}

\abstract{We study the prospects for an independent determination of $V_{cb}$ from bottom-tagged charged-current deep-inelastic scattering at future electron-proton colliders: the EIC, LHeC, and FCC-eh.
The transition $e^- c \to \nu_e b$ probes $|V_{cb}|^2$ through a process with different theoretical and experimental uncertainties from semileptonic $B$ decays.
We fit ratios of $b$-tagged charged-current event counts over flavour-inclusive ones in kinematic bins, including correlated parton-distribution and flavour-tag calibration uncertainties, and floating the other CKM parameters.
With projected parton distributions and tagging performance motivated by existing studies, we find a potential relative sensitivity in $|V_{cb}|$ of about $0.8\%$ at the LHeC with $1\,\rm{ab}^{-1}$ of integrated luminosity and $0.7\%$ at FCC-eh with $2\,\rm{ab}^{-1}$.
The corresponding EIC sensitivity is instead at the level of $20 - 30 \%$, depending on the configuration.
The LHeC could thus provide a high-energy determination competitive with low-energy ones, but with substantially different dominant systematic uncertainties, on the timescale of its proposed operation between the HL-LHC and FCC-ee.
We identify precise flavour-tag calibration and control of the kinematic response as the main requirements for realising this potential sensitivity.}

\begin{document}

\maketitle

\section{Introduction}
\label{sec:introduction}

The Cabibbo-Kobayashi-Maskawa (CKM) matrix element $V_{cb}$ is a central input to flavour physics, both in the Standard Model (SM) and beyond.
Besides governing the dominant semileptonic decays of $B$ mesons, it enters SM predictions for rare kaon and $B$ decays, neutral-meson mixing, and CP-violating processes.
Its uncertainty often limits the sensitivity of these measurements to new physics.
The most precise determinations use exclusive $B\to D^{(*)} \ell \nu$ (where $\ell = e, \mu$) or inclusive $B\to X_c \ell \nu$ decays.
The updated 2026 PDG averages for these two classes of measurements are~\cite{ParticleDataGroup:2026mpi}
\begin{equation}
    \Vcb_{\rm excl}=(39.5\pm0.5)\times10^{-3},\qquad
    \Vcb_{\rm incl}=(42.0\pm0.5)\times10^{-3}\,,
    \label{eq:low-energy-vcb}
\end{equation}
showing a longstanding and significant tension between the two.
Further improvements in the precision from $B$ decays require better control of experimental and theoretical systematic uncertainties.
A determination with different systematics is thus particularly valuable.

High-energy processes offer several possibilities.
The $W^+\to c\bar b$ decay directly probes the same charged-current coupling, provided that the two quark flavours can be identified.
ATLAS has recently measured $\Vcb=(50^{+11}_{-14})\times10^{-3}$ using $W$ bosons from $t \bar{t}$ production \cite{ATLAS:2026gkw}, see also Ref.~\cite{Zhao:2025rci}.
At FCC-ee, the very large $W$ sample, the clean environment, and the possibility to calibrate jet tagging efficiencies with very high precision at the $Z$ pole could improve this precision substantially~\cite{Charles:2020dfl,Marzocca:2024mkc}.
Ref.~\cite{Marzocca:2024mkc} finds a projected relative sensitivity of $0.52\%$ with 1\% tag-calibration uncertainties, improving to $0.16\%$ when these are reduced to 0.1\%.
A different opportunity is provided by the intense neutrino beam from a high-energy muon collider, for which deep-inelastic-scattering (DIS) measurements could also reach the few-per-mille level~\cite{Marzocca:2025inb}.

In this work we study the prospects for a $V_{cb}$ measurement through charged-current (CC) DIS with a $b$-tagged jet at electron-proton colliders,
\begin{equation}
    e^- + p \to \slashed{E}_T + b\,\text{-jet} + X.
    \label{eq:process}
\end{equation}
At the Born level, the signal is the partonic transition $e^- c \to \nu_e b$, whose cross section is directly proportional to $\Vcb^2$.
The main challenge is to distinguish this CKM-suppressed process from the much larger light-flavour and anticharm CC samples.
Bottom tagging provides the main background rejection, while the different distributions in the DIS variables provide further discrimination.

We consider three future colliders: the Electron-Ion Collider (EIC)~\cite{AbdulKhalek:2021gbh}, planned to start taking physics data in 2037, the Large Hadron-electron Collider (LHeC)~\cite{LHeC:2020van,Ahmadova:2025vzd}, proposed as a bridge project for CERN between the end of HL-LHC and the start of FCC-ee, and FCC-eh~\cite{FCC:2018byv}, which could run concurrently with the hadron-hadron collider.
All three machines would improve our knowledge of proton parton distribution functions (PDFs), including the charm distribution entering our signal, and offer a rich physics program for SM precision measurements and searches for new physics.

Our aim is to establish the $\Vcb$ precision potential of each experiment and the performance required to realise that potential. The calculation is done at the Born level and uses flat tagging responses.
In our statistical analysis, we propagate PDF, CKM (except for the target $\Vcb$), and tagging uncertainties. We assess the impact of transverse-momentum thresholds, of the assumptions on tagging efficiencies and their uncertainties, and of unknown shape uncertainties.
No external $\Vcb$ prior or CKM-unitarity relation is imposed in the fit.
Our sensitivity forecast serves as motivation for more detailed studies, in which residual theory uncertainties and detector effects could be established on firmer grounds.

Section~\ref{sec:experiments} introduces the collider scenarios and flavour tagging assumptions.
Section~\ref{sec:processes} describes the CC signal and backgrounds and the PDF sets used. Section~\ref{sec:observables} introduces the observables we use and the construction of the likelihood.
Our main results are presented in Section~\ref{sec:results}, along with their dependence on the main analysis inputs. Finally, we conclude in Section~\ref{sec:conclusions}.
The Born cross sections and massive kinematics are collected in Appendix~\ref{app:cross-sections} while details on the statistical analysis are given in Appendix~\ref{app:statistics}.

\section{Experimental scenarios}
\label{sec:experiments}

\begin{table}[t]
 \centering
 \begin{tabular}{lrrrr}
 \toprule
 Collider & $E_e$ [GeV] & $E_p$ [TeV] & $\sqrt{s}$ [TeV] & $\mathcal L$ [fb$^{-1}$] \\
 \midrule
 EIC & 10 & 0.275 & 0.105 & 200 \\
 EIC & 18 & 0.275 & 0.141 & 46 \\
 \midrule
 LHeC & 50 & 7 & 1.18 & 1000 \\
 FCC-eh & 60 & 42.5 & 3.19 & 1000 / 2000 \\
 \bottomrule
 \end{tabular}
 \caption{Adopted electron-proton collider scenarios.}
 \label{tab:colliders}
\end{table}

The beam energies and integrated luminosities we use are listed in Table~\ref{tab:colliders}.
Each row is a separate projection and different scenarios are not combined.

For the EIC \cite{AbdulKhalek:2021gbh} we use the high-divergence machine parameters and the two highest electron energies, 10 and 18 GeV, with 275 GeV protons, for total integrated luminosities of $200\,\mathrm{fb}^{-1}$ and $46\,\mathrm{fb}^{-1}$, respectively.\footnote{The collider parameters are taken from the Master Parameter Table at  \href{https://eic.jlab.org/Parameters/index.html}{\tt https://eic.jlab.org/Parameters/index.html}. To obtain the integrated luminosities we assume $10^7$ seconds per nominal year and stage durations of two and three years for the 10 and 18 GeV electron runs, respectively, as discussed for the pre-TDR planning scenarios: \href{https://indico.bnl.gov/event/28486/}{\tt https://indico.bnl.gov/event/28486/}.}

The LHeC would collide the 7 TeV HL-LHC proton beam with electrons from an energy-recovery linac \cite{LHeC:2020van}.
The recent bridge proposal uses a 50 GeV electron beam and envisages approximately $1\,\mathrm{ab}^{-1}$ in six years of operation after the HL-LHC~\cite{Ahmadova:2025vzd}.
For FCC-eh, 60 GeV electrons collide with the 42.5 TeV proton beam of the 85 TeV FCC-hh design baseline~\cite{FCC:2018byv,FCC:2025uan}. The total integrated luminosity for FCC-eh is expected at $2\,\mathrm{ab}^{-1}$. We also consider a lower exposure case of $1\,\mathrm{ab}^{-1}$.

All three colliders envisage the possibility of colliding polarised electrons.
We define the electron polarisation as $P_e=(N_R-N_L)/(N_R+N_L)$.
For massless electrons, the CC DIS cross section satisfies
\begin{equation}
 \sigma_{\rm CC}(P_e)=(1-P_e)\sigma_{\rm CC}(0).
 \label{eq:polarisation}
\end{equation}
We compare unpolarised beams with polarised electrons, with $P_e=-0.7$ at the EIC \cite{AbdulKhalek:2021gbh} and $P_e=-0.8$ at the LHeC and FCC-eh \cite{LHeC:2020van,Andre:2025iar,FCC:2018byv}.

\subsection{Jet flavour tagging}
\label{sec:tagging}

Recent improvements in flavour tagging motivate the use of bottom-tagged DIS for a CKM measurement.
Bottom-tagged final states at high-energy electron-proton colliders have been studied for Higgs to $b \bar b$ \cite{Han:2009pe,Behera:2022gnr,Haghighat:2023bgi} and single-top production \cite{Dutta:2013mva,Gao:2021plf}.
These studies address different jet kinematics and event selections.
A dedicated study of tagging performance and calibration, with modern jet tagging techniques, for the CC DIS sample considered here is still needed.
We therefore adopt reasonable performances from related studies, and examine later the dependence of our results on the tagging efficiencies.

For the EIC, Ref.~\cite{Boettcher:2024yax} studies a secondary-vertex event tag in predominantly neutral-current DIS.
It reports a typical bottom efficiency of $0.30$--$0.40$, charm mistags of approximately $0.01$ and light mistags of approximately $10^{-4}$. We take the central values and assume that these efficiencies can be transferred, to a good approximation, to our process of interest.
Related EIC studies have considered CC charm tagging as a probe of the strange PDF~\cite{Arratia:2020azl}, bottom-tagged NC DIS to measure the $Z b \bar b$ coupling \cite{Yan:2021htf,Li:2021uww}, and bottom-tagged final states as probes of new flavour-changing interactions~\cite{Bar-Shalom:2026fou}. 

For the higher energy LHeC and FCC-eh colliders we use tagging efficiencies motivated by recent studies performed for LHC experiments. Specifically, we consider the recent CMS `Unified Particle Transformer algorithm', called UParT v2~\cite{CMS-DP-2025-081}.
The bottom efficiency is 0.60, with approximate charm and light quark mistag probabilities of $1.6 \times 10^{-3}$ and $6\times10^{-5}$, respectively.\footnote{The latter is obtained by a modest extrapolation below the displayed $10^{-4}$ range.}
We compare it with results from the CMS ParticleNet B4 algorithm \cite{CMS:2025dsh} and a hypothetical improved response in Section~\ref{sec:calibration}.

Tagging efficiencies $\epsilon_q$ are defined as probabilities of tagging (or mistagging) a jet of true flavour $q$ as a $b$-jet. The label `light' indicates the $u,d,s$ flavours.
To account for calibration uncertainties in the tagging efficiencies, for each flavour $f=b,c,\mathrm{light}$ we introduce one independent Gaussian calibration pull, shared across bins and all tagged components of that flavour,
\begin{equation}
 \epsilon_f(a_f)=\epsilon_f^0 e^{\delta_f a_f},\qquad a_f\sim\mathcal N(0,1),
 \label{eq:tag-response}
\end{equation}
where $\epsilon_f^0$ is the central value, $\delta_f$ the corresponding relative uncertainty, and $a_f$ the associated standard Gaussian nuisance parameter.
The baseline relative calibration uncertainties are $(5,10,10)\%$ for the EIC and $(1,5,5)\%$ for the LHeC and FCC-eh.
We denote these triplets for each collider as $C_0$ in the following.
The nominal probabilities and relative calibration uncertainties assumed for our baseline analyses are collected in Table~\ref{tab:tagging}. 

\begin{table}[t]
 \centering
 \begin{tabular}{l|rrr|rrr}
 \toprule
 & $\epsilon_b$ & $\epsilon_c$ & $\epsilon_{\rm light}$ & $\delta_b$ & $\delta_c$ & $\delta_{\rm light}$ \\
 \midrule
 EIC & 0.35 & 0.01 & $10^{-4}$ & 5\% & 10\% & 10\% \\
 LHeC/FCC-eh & 0.60 & 0.0016 & $6\times10^{-5}$ & 1\% & 5\% & 5\% \\
 \bottomrule
 \end{tabular}
 \caption{Baseline $b$-jet tagging probabilities and relative one-standard-deviation calibration uncertainties.}
 \label{tab:tagging}
\end{table}

The impact of different choices for these parameters on the $V_{cb}$ sensitivity is explored in Section~\ref{sec:calibration}.
Our simplifying assumption is to take the tagging efficiencies as constant in the whole kinematical range. In Section~\ref{sec:shape} we discuss how our results depend on uncertainties in their kinematic dependence.
The adopted tagging probabilities describe the response to jets of a specified final-state flavour. Additional bottom production from QCD radiation requires a consistent treatment of the heavy-flavour production rate, jet-flavour definition and tagging response, which is beyond the present Born-level study.

Another strategy to describe flavourful hadronic production in DIS is to study Semi-Inclusive DIS (SIDIS), employing in our case $B$ hadron fragmentation functions (FFs). This method was adopted for example in Ref.~\cite{Marzocca:2025inb}, for the CKM study in DIS with the muon collider neutrino beam. In Section~\ref{sec:sidis} we adopt this alternative flavour tagging description and compare its performance with the jet flavour tagging one.

\section{Signal and background processes}
\label{sec:processes}

\begin{figure}[t]
 \centering
\begin{tikzpicture}
\begin{feynman}
  \vertex (e)  at (0,3)    {$e^-$};
  \vertex (a)  at (3,3);
  \vertex (nu) at (6,3)    {$\nu_e$};
  \vertex (p)  at (0,0)    {$p$};
  \vertex (b)  at (6,0.8)  {$b$};
  \vertex (X)  at (5,-0.7) {$X$};

  \vertex[blob,pattern=none,fill=gray!20,minimum size=5mm]
    (P) at (1.5,0) {};
  \vertex[dot,label=above left:{$V_{cb}$}]
    (v) at (3,0.8) {};

  \diagram* {
    (e) -- [fermion] (a) -- [fermion] (nu),
    (a) -- [boson,edge label={$W^-$}] (v),
    (p) -- [plain] (P),
    (P) -- [fermion,edge label'={$c$}] (v) -- [fermion] (b),
    (P) -- [plain] (X),
  };
\end{feynman}
\end{tikzpicture}
\caption{The charm-initiated Born contribution to $e^-p\to\nu_e b+X$.
The charged-current vertex probes $V_{cb}$.}
 \label{fig:process}
\end{figure}
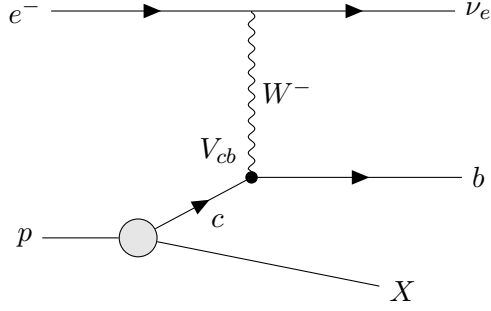

At Born level the bottom final state receives two contributions,
\begin{equation}
 \mathrm{CC}_{b}: \quad e^-c\to\nu_e b\quad (\sigma_{\rm CC}^{cb}\propto\Vcb^2),\qquad
 e^-u\to\nu_e b\quad (\sigma_{\rm CC}^{ub}\propto\Vub^2).
 \label{eq:signal-channels}
\end{equation}
For the up-quark channel, the smaller CKM factor is partly compensated by the larger valence up PDF.
The CC DIS mistag backgrounds are
\begin{equation}
\begin{aligned}
 \mathrm{CC}_{\bar c}:&\quad e^-(\bar d,\bar s,\bar b)\to\nu_e\bar c,\\
 \mathrm{CC}_{\rm light}:&\quad e^-(u,c)\to\nu_e(d,s),\qquad
 e^-(\bar d,\bar s,\bar b)\to\nu_e\bar u.
\end{aligned}
 \label{eq:background-channels}
\end{equation}
The CKM factors are kept explicitly and are varied in the fit with central values and prior uncertainties taken from the PDG averages \cite{ParticleDataGroup:2026mpi}:
\begin{equation}
\begin{aligned}
 |V_{ud}|&=0.97367\pm0.00032,& |V_{us}|&=0.22431\pm0.00085, & |V_{ub}|&=(3.89\pm0.16)\times 10^{-3}\\
 |V_{cd}|&=0.2247\pm0.0022,& |V_{cs}|&=0.969\pm0.005, & |V_{cb}|&=(4.07\pm0.13)\times 10^{-2}, \\
 |V_{td}|&=(8.6\pm0.2)\times 10^{-3},&|V_{ts}|&=(4.15\pm0.09)\times 10^{-2},&|V_{tb}|&=1.010\pm0.027.
\end{aligned}
 \label{eq:ckm-inputs}
\end{equation}
The last row is used for the single-top background discussed below. For $\Vcb$ we only use the central value as reference value, $\Vcb_{\rm ref}=0.0407$, to define the expected data, while its present uncertainty is not used as a prior.
There is no upper bound of one on the direct $|V_{tb}|$ input and no unitarity constraint on the CKM matrix elements.
In our baseline analysis we neglect the background from $b$ quark production in neutral-current (NC) DIS, where the electron goes undetected. Our estimate of the possible impact of this contribution is discussed in Section~\ref{sec:nc}.

For $e^-(k)p(P)\to\nu_e(k')+X$ the DIS variables are defined as
\begin{equation}
 s=(P+k)^2 \simeq 4E_eE_p \,,
 \quad q=k-k'\,, 
 \quad Q^2=-q^2\,, 
 \quad x=\frac{Q^2}{2P\cdot q}\,,
 \quad y=\frac{Q^2}{xs}\,.
 \label{eq:dis-variables}
\end{equation}
The full $2 \to 2$ kinematics and DIS cross sections are collected in Appendix~\ref{app:cross-sections}.
Antiquark-initiated channels carry a $(1-y)^2$ weight, whereas the quark-initiated signal does not.
Measuring the $y$ dependence therefore helps reject anticharm.
Additional separation of signal from background can be derived from the different $x$ dependence of the charm and up PDFs.
We calculate the $c \to b$ and $u \to b$ processes with an on-shell outgoing bottom mass $m_b=4.75\,\mathrm{GeV}$ and massless incoming quarks.
The bottom mass changes the PDF argument to $\xi=x(1+m_b^2/Q^2)$. Other kinematical variables are also modified, with expressions given in Appendix~\ref{app:cross-sections}.
The remaining non-top CC processes are evaluated in the massless limit.

The DIS cuts common to the three colliders are $Q^2>100\,\mathrm{GeV}^2$ and $W^2>10\,\mathrm{GeV}^2$.
The remaining cuts are
\begin{equation}
\begin{array}{rlll}
 \text{EIC:}&\quad 0.01<y<0.95,& \quad |\eta|<3.5,& \quad \pt=\pt^{\rm miss}>5\,\mathrm{GeV},\\[2pt]
 \text{LHeC, FCC-eh:}&\quad 0.05<y<0.9,& \quad |\eta|<3.6,& \quad \pt=\pt^{\rm miss}>10\,\mathrm{GeV}.
\end{array}
 \label{eq:cuts}
\end{equation}
Here $\pt$ and $\eta$ refer to the outgoing parton.
The EIC pseudorapidity cut is taken as the tracker acceptance in the ePIC pre-TDR report~\cite{dalla_torre_2026_18271602}.
The LHeC detector design motivates tracking through $|\eta|<3.6$~\cite{Andre:2022xeh}, which we also assume for FCC-eh.
The dependence of the $\Vcb$ precision on the minimum $p_T$ threshold is studied in Section~\ref{sec:threshold}.

\subsection{PDFs}
\label{sec:pdfs}

The signal measures the product of $\Vcb^2$ and the charm PDF.
A precise determination thus requires both external PDF information and the flavour discrimination provided by jet tagging and the kinematic distributions.
Since electron-proton colliders are expected to improve PDF determinations, we include future PDF constraints using published projections.

For the EIC we use the \emph{pessimistic} projection of Ref.~\cite{Khalek:2021ulf}, based on NNPDF3.1. We checked that using the \emph{optimistic} projection does not change our results substantially.
The set we employ is a Hessian representation of the original 100-replica Monte Carlo ensemble, with 50 symmetric uncertainty directions and one central member, obtained using the \texttt{mc2hessian} method~\cite{Carrazza:2015aoa}.
The source combines inclusive proton NC/CC and deuteron NC pseudodata from EIC with existing measurements.

For the LHeC we use the LHeC PDF projection set \texttt{PDF4LHC15\_nnlo\_lhec}, with 100 symmetric Hessian directions~\cite{AbdulKhalek:2019mps}. This projection also includes charm-tagged NC and CC DIS and bottom-tagged NC DIS, but no bottom-tagged CC data. The tagged CC observable considered here is therefore not used to obtain the projected PDFs.
In the absence of a dedicated projection, the same set is also used for FCC-eh.

The final $V_{cb}$ sensitivities are compared with those obtained with the present-data set PDF4LHC21~\cite{PDF4LHCWorkingGroup:2022cjn}, in its 40-direction five-flavour Hessian version.
All three PDF inputs have NNLO evolution and $\alpha_s(M_Z)\simeq0.118$.

\begin{figure}[t]
 \centering
 \includegraphics[width=0.9\textwidth]{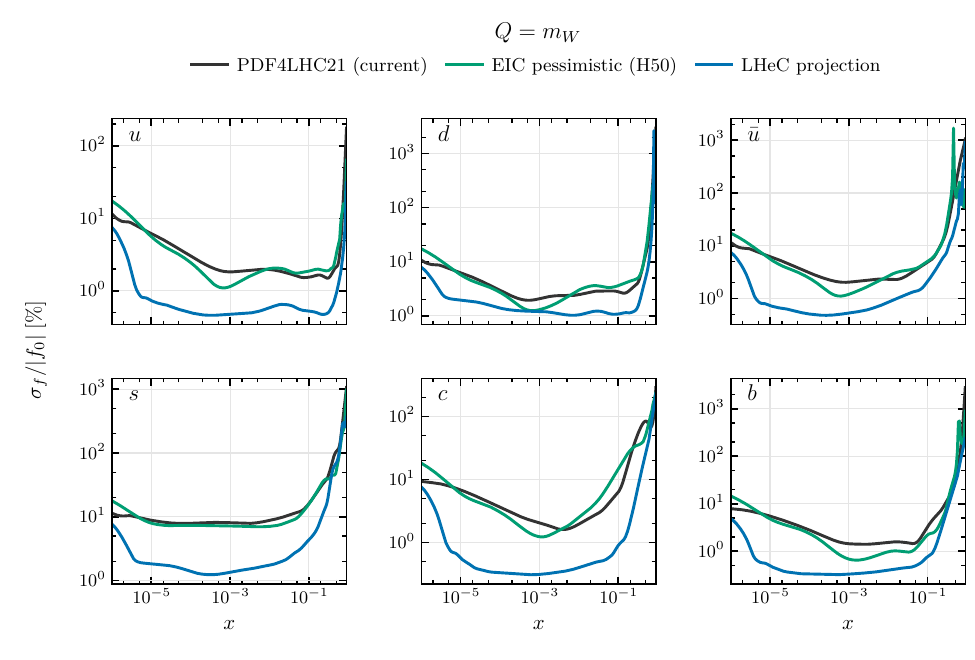}
 \vspace{-5mm}
 \caption{Relative $1\sigma$ PDF uncertainties at $Q=M_W$ for the representative partons.
The curves show PDF4LHC21 (black), the pessimistic EIC projection (green) and the LHeC projection used also for FCC-eh (blue).}
 \label{fig:pdfs}
\end{figure}

For each parton flavour $F$, the symmetric Hessian sets contain $N_H + 1$ functions $f_F^{(k)}(x, Q^2)$, where $k=0,1,\ldots,N_H$ and $k=0$ denotes the central member. The PDF nuisance dependence is parametrised as
\begin{equation}
    f_F(x, Q^2; \nu) = f_F^{(0)} + \sum_{k=1}^{N_H} \nu_k \left[ f_F^{(k)} - f_F^{(0)}\right] \, .
\end{equation}
This dependence is common across flavours, bins, and processes, including top production where present.

The correlated one-standard-deviation PDF uncertainty can be computed as
\begin{equation}
 (\Delta f_F)^2=\sum_{k=1}^{N_H}\left(f_F^{(k)}-f_F^{(0)}\right)^2 \, .
 \label{eq:hessian}
\end{equation}
Figure~\ref{fig:pdfs} compares the uncertainty of the relevant flavours at $Q=M_W$ for the three sets.

\subsection{Top-quark background}
\label{sec:top}

Single-antitop production is kinematically accessible at the LHeC and FCC-eh and has no $V_{cb}$ suppression in its dominant channel, 
\begin{equation}
 \mathrm{CC}_{\bar t}: \quad e^- \bar b \to \nu_e \bar t \,, \quad \bar t\to W^-\bar b \, .
\end{equation}
These events can enter the $b$-tagged CC category if only one jet is within the acceptance cuts and the other decay products evade the lepton veto or jet requirements.
We calculate the selected contribution using ten million partonic Born-level events generated with MadGraph5\_aMC@NLO~\cite{Alwall:2014hca}.

The prediction includes $e^-\bar q\to\nu_e\bar t$ and $\bar t\to W^-\bar q'$ for $q,q'=d,s,b$.
All hadronic and leptonic $W$ boson decays are included, and the $\tau$ lepton is not decayed.
The stored events are reweighted for PDFs as well as production and decay CKM factors.
All non-top quarks are treated as massless in this simulation.

We cluster jets with anti-$k_T$, $R=0.4$, using FastJet~\cite{Cacciari:2008gp,Cacciari:2011ma}.
We require exactly one fiducial jet with $\pt>10\,\mathrm{GeV}$ and $|\eta|<3.6$, together with the event-level DIS and missing-momentum cuts in Eq.~\eqref{eq:cuts}.
We also require the jet and missing transverse momenta to balance within 5 GeV:
\begin{equation}
 \left|\boldsymbol p_T^{\rm miss}+\boldsymbol p_{T,j}\right|<5\,\mathrm{GeV} \, .
 \label{eq:balance}
\end{equation}
An electron, muon or stable tau vetoes the event if it has $\pt>10\,\mathrm{GeV}$, $|\eta|<3.6$ and $\Delta R_{\eta\phi}>0.4$ from every fiducial jet.
Identification is assumed perfect within this window.
A selected jet is labelled bottom if it contains a bottom parton, otherwise charm if it contains charm, and light otherwise.
This includes events in which the accepted jet originates from the $W$ and is mistagged.
The Born-level non-top CC DIS final states automatically satisfy these additional requirements.

For top events we reconstruct the CC DIS kinematic variables from the full visible final state with the Jacquet–Blondel method \cite{Amaldi:1979qp},
\begin{equation}
 y_{\rm vis}=\frac{\sum_a(E_a-p_{z,a})}{2E_e} \, , \qquad
 Q^2_{\rm vis}=\frac{|\boldsymbol p_T^{\rm miss}|^2}{1-y_{\rm vis}} \, , \qquad
 x_{\rm vis}=\frac{Q^2_{\rm vis}}{s y_{\rm vis}} \, .
 \label{eq:visible}
\end{equation}
The sum runs over all final-state particles other than neutrinos, including charged leptons and particles outside the identification acceptance.
The missing transverse momentum is the vector sum of all final neutrinos.
The event-level DIS cuts and bin assignments use these visible-state estimators.

At $P_e=0$, the selected and tagged top cross sections are approximately $3.0$ fb at the LHeC and $27$ fb at FCC-eh for our baseline tagger.
Direct leptonic $W$-decay channels dominate after the exclusive selection.
The top background shares PDF, CKM and tagging nuisance parameters with the other channels and is assigned an additional 10\% normalisation uncertainty.
With our top background modelling and selection, increasing this uncertainty to 30\% has almost no impact on the projected $\Vcb$ precision.

\section{Observables}
\label{sec:observables}

\begin{table}[t]
 \centering
 \small
 \begin{tabular}{lrrrrr}
 \toprule
 Collider, $\mathcal L$ & $c \to b$ & $u \to b$ & $\bar q \to \bar c$ & Light & Top \\
 \midrule
 EIC $10\times275$, $200\,\mathrm{fb}^{-1}$ & 72.3 & 12.9 & 594 & 277 & --- \\
 EIC $18\times275$, $46\,\mathrm{fb}^{-1}$ & 33.6 & 4.85 & 306 & 105 & --- \\
 \midrule
 LHeC, $1\,\mathrm{ab}^{-1}$ & 39\,328 & 945 & 36\,793 & 10\,350 & 3\,018 \\
 FCC-eh, $1\,\mathrm{ab}^{-1}$ & 109\,913 & 1\,599 & 101\,551 & 21\,405 & 26\,908 \\
 FCC-eh, $2\,\mathrm{ab}^{-1}$ & 219\,826 & 3\,198 & 203\,103 & 42\,810 & 53\,816 \\
 \bottomrule
 \end{tabular}
 \caption{Expected tagged event counts in each final-state channel for the cuts in Eq.~\eqref{eq:cuts}, the responses in Table~\ref{tab:tagging}, projected PDFs and $P_e=0$. The top selection is defined in Section~\ref{sec:top}.}
 \label{tab:yields}
\end{table}

Table~\ref{tab:yields} shows the tagged event yields for each final-state channel, accounting for the tag probabilities of Table~\ref{tab:tagging} and the selection discussed in the previous section, and for unpolarised electrons. Event counts for polarised electrons can be obtained by multiplying these numbers by $(1-P_e)$.
The EIC has tens of selected $cb$ events, against a substantially larger mistag background, mostly from anticharm.
At the LHeC and FCC-eh, the signal reaches tens to hundreds of thousands of events and is comparable to the anticharm contribution, which again is by far the largest background source.

\begin{figure}[t]
 \centering
 \includegraphics[width=0.9\textwidth]{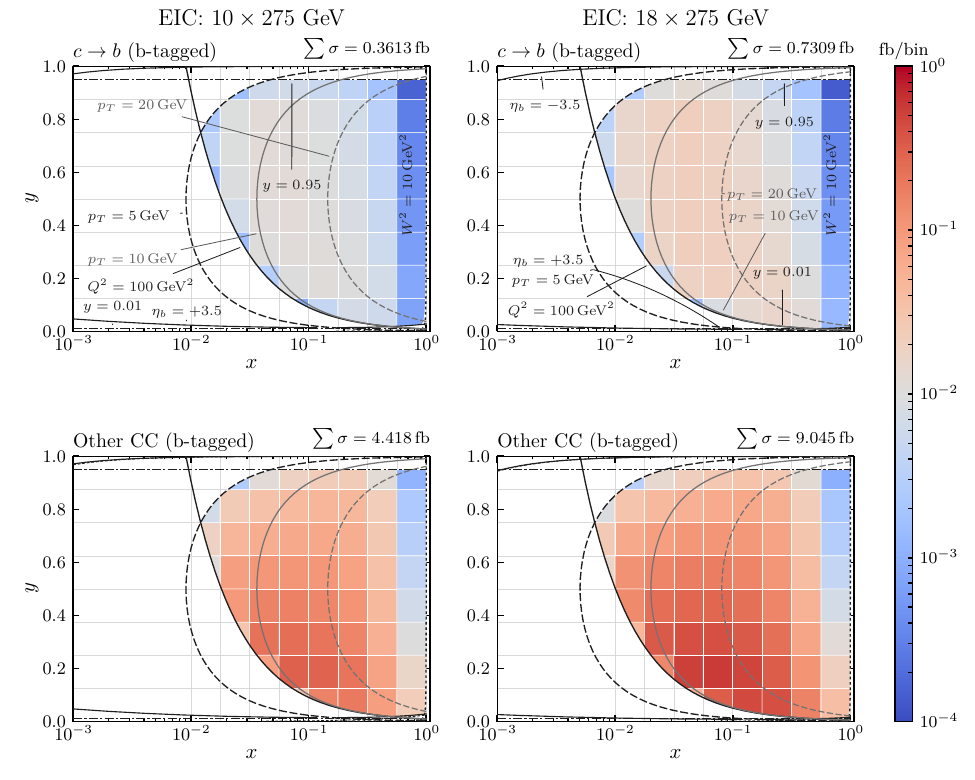}\vspace{-0.3cm}
 \caption{Bin-integrated cross sections, in fb, for the EIC at $P_e=0$ for $10\times275$ GeV (left) and $18\times275$ GeV (right). Rows show the $c \to b$ signal (top) and the other background CC channels (bottom), after $b$-tagging. Black and grey curves indicate the selection boundaries, as indicated in the figures.}
 \label{fig:eic-rates}
\end{figure}

\begin{figure}[t]
 \centering
 \includegraphics[width=0.9\textwidth]{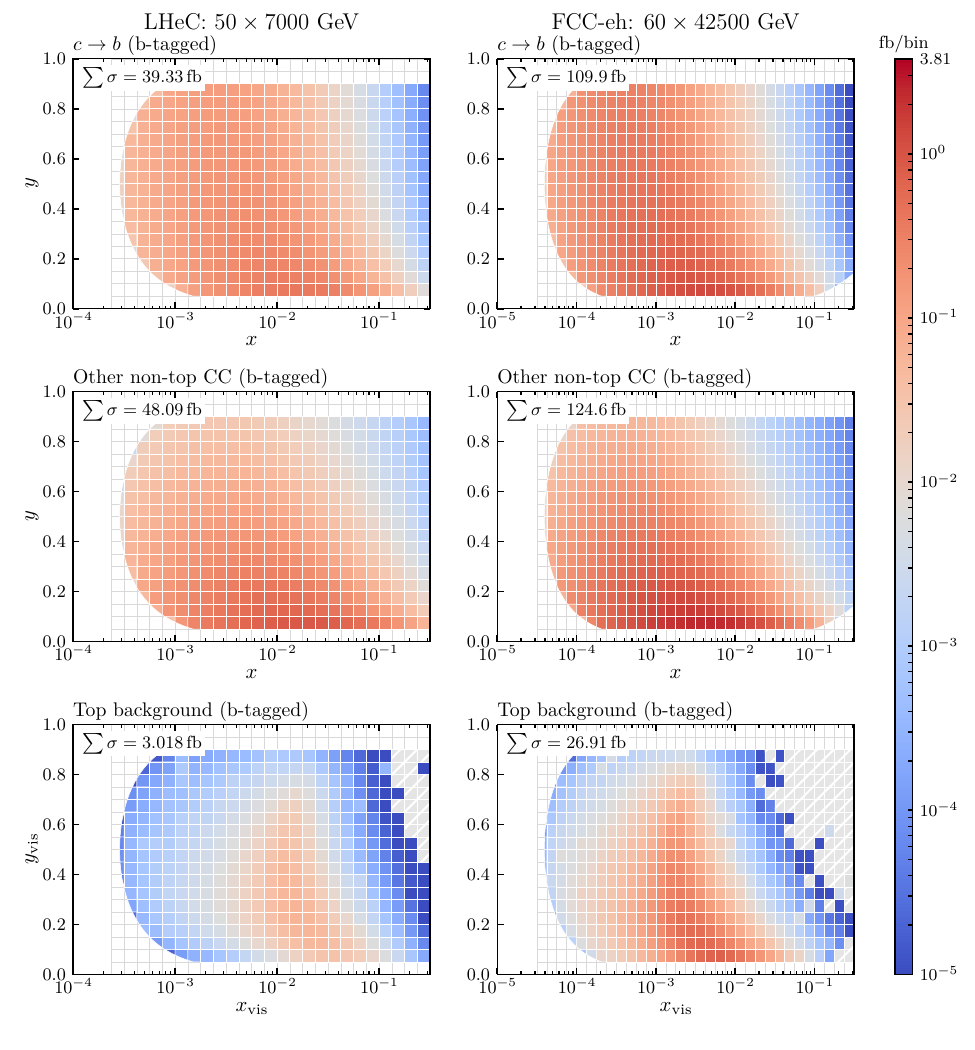}\vspace{-0.5cm}
 \caption{Bin-integrated cross sections, in fb, for the LHeC (left) and FCC-eh (right) at $P_e=0$. Rows show the $c \to b$ signal (top), the other non-top CC channels (central), and the top background passing the selection (bottom), after $b$-tagging.}
 \label{fig:high-energy-rates}
\end{figure}

We integrate the differential DIS cross section in rectangular $(x,y)$ bins.
For the EIC we divide $x \in [10^{-2}, 1]$ into 8 logarithmically spaced bins, plus one for $x < 10^{-2}$, and $y\in [0, 1]$ into 8 equally spaced bins.
For the LHeC and FCC-eh we divide each decade of $x$ into 8 logarithmically spaced bins, starting at the smallest $x$ value with a non-zero cross section and ending at $x \approx 0.3$, and $y\in [0, 1]$ into 20 equally spaced bins.
These bin sizes were chosen to retain kinematic discrimination while avoiding potential issues with detector resolution and smearing, which are not accounted for in this work.
The dependence of the $\Vcb$ precision on bin size is studied in Section~\ref{sec:results}.

Figures~\ref{fig:eic-rates} and \ref{fig:high-energy-rates} show the cross sections integrated in each $(x,y)$ bin for unpolarised electrons, for the $cb$ signal, the other non-top tagged CC channels and, for LHeC and FCC-eh, the top-induced background.
In the top panels of Figure~\ref{fig:eic-rates} we label explicitly the impact of each acceptance cut. 

For each kinematic bin $i$ we define the ratio of the $b$-tagged event count to the corresponding flavour-inclusive selected CC DIS count:
\begin{equation}
 R_i(\mu_{cb},\bm\theta)=\frac{N_i^{b\text{-tag}}(\mu_{cb},\bm\theta)}{N_i^{\rm incl}(\mu_{cb},\bm\theta)} \, ,
 \label{eq:ratio}
\end{equation}
where $\mu_{cb}=\Vcb^2/\Vcb^2_{\rm ref}$ is our signal strength and $\bm\theta = \left( \nu_I, V_\alpha - V_\alpha^{\rm ref} \right)$ contains all the unit-Gaussian nuisance parameters $\nu_I$ and the floating CKM matrix elements $V_\alpha = \{ |V_{ij}| \}$ (except for $\Vcb$, for which we set no prior).
The inclusive sample has the same cuts and binning as the tagged one, with only the flavour-tag requirement removed.
Both predictions retain their full PDF and CKM dependence.
Selected top events contribute before tagging to the denominator and after tagging to the numerator.
Flavour-tag calibration and the extra mistag-response modes affect only the numerator.

Compared to absolute event counts, this ratio is a more robust observable as corrections that act as common multiplicative factors cancel.
Residual effects from flavour-dependent PDF convolutions, quark masses and the exclusive selection require a dedicated calculation. 
In the ratio, the CC polarisation factor in Eq.~\eqref{eq:polarisation} also cancels, although polarisation increases the event sample and hence the statistical precision.

For each bin, the independent Poisson likelihoods for the tagged, $N_i^{b\text{-tag}}$, and untagged, $N_i^{\rm untag} = N_i^{\rm incl} - N_i^{b\text{-tag}}$, counts can be factorised into a Poisson likelihood for the inclusive count and a conditional binomial likelihood for the tagged ratio $R_i$. We keep only the conditional term.
As a result, our test statistic can be written as
\begin{equation}
 \Delta \chi^2(\mu_{cb}, \bm \theta) = 2 \sum_i N_{i,0}^\text{incl} \left[R_{i,0}\log\frac{R_{i,0}}{R_i}+(1-R_{i,0})\log\frac{1 - R_{i,0}}{1-R_i}\right]
 + \bm\theta^{\mathsf T} \Sigma_{\rm aux}^{-1} \bm\theta \, ,
 \label{eq:chi2}
\end{equation}
where $R_{i,0} \equiv R_i(1,\bm 0)$ and $\Sigma_{\rm aux}$ is the auxiliary covariance matrix of the nuisance parameters. By construction, $\Delta \chi^2$ vanishes at the reference point \mbox{$(\mu_{cb}, \bm \theta)_{\rm ref} = (1, \bm 0)$}.
From this, we obtain the profiled local Fisher uncertainty $\delta \Vcb / \Vcb$.
Details on the statistics can be found in Appendix~\ref{app:statistics}.

\section{Results}
\label{sec:results}

\begin{table}[t]
 \centering
 \small
 \begin{tabular}{l|r|r}
 \toprule
 \multicolumn{3}{c}{$\delta|V_{cb}|/|V_{cb}|\;[\%]$} \\
 \midrule
 Scenario & $P_e=0$ & $P_e<0$ \\
 \midrule
 EIC $10\times275$, $200\,\mathrm{fb}^{-1}$ & 26 & 22 \\
 EIC $18\times275$, $46\,\mathrm{fb}^{-1}$ & 32 & 26 \\
 LHeC, $1\,\mathrm{ab}^{-1}$ & 0.89 & 0.79 \\
 FCC-eh, $1\,\mathrm{ab}^{-1}$ & 0.75 & 0.70 \\
 FCC-eh, $2\,\mathrm{ab}^{-1}$ & 0.70 & 0.67 \\
 \bottomrule
 \end{tabular}
 \caption{Projected $1\sigma$ relative uncertainty on $\Vcb$ [\%] from our baseline fits, including CKM and projected PDF uncertainties, top background with 10\% normalisation width, and tagging efficiencies and calibration as per Table~\ref{tab:tagging}. The polarised scenarios use $P_e=-0.7$ for the EIC and $P_e=-0.8$ for LHeC and FCC-eh.}
 \label{tab:vcb-results}
\end{table}

Our main results for the projected uncertainty on $\Vcb$ from the baseline fits are collected in Table~\ref{tab:vcb-results}. The numbers shown are relative Fisher uncertainties, in percent.
The baseline $\Vcb$ uncertainty is obtained after profiling over projected PDF uncertainties, uncertainties in all other CKM elements, a 10\% normalisation uncertainty of the top contribution (for LHeC and FCC-eh), and the flavour tagging calibration uncertainties as shown in Table~\ref{tab:tagging}.

The projected precision at the EIC ranges from 22\% to 32\%, depending on the scenario and polarisation.
The $10\times 275$ GeV stage performs better than the higher-energy stage because of its larger assumed luminosity.
Although these scenarios are not competitive with the present $V_{cb}$ precision, they could test the coupling in a different process and develop the required $b$-tagging methods, which could also prove useful for other measurements. We note that in all EIC scenarios, the $\Vcb = 0$ point would be compatible within the $3\sigma$ level, computed using profiled $\Delta\chi^2$ intervals.

On the other hand, all high-energy $e^-p$ colliders could reach sub-percent precision.
Interestingly, our sensitivity projection for the LHeC is $\delta\Vcb/\Vcb=0.89\%$ for unpolarised electrons and $0.79\%$ for $P_e=-0.8$. 
FCC-eh with $2\,\mathrm{ab}^{-1}$ ($1\,\mathrm{ab}^{-1}$) and polarised electrons could improve this to $0.67\%$ ($0.70\%$).
The precision achievable at both LHeC and FCC-eh is competitive with the one from semileptonic $B$ decays.
Such a measurement would be extremely valuable for understanding the current tension between the inclusive and exclusive decays, providing an independent test with different dominant systematic uncertainties.

Our baseline fits use projected PDFs. To see their impact we also performed fits using the current set, PDF4LHC21. With this set, the projected uncertainty for polarised electrons increases from 0.79\% to 1.0\% at the LHeC and from 0.67\% to 0.83\% at FCC-eh with $2\,\mathrm{ab}^{-1}$.
However, we believe that the projected PDF sets offer a much more plausible description of the PDF precision that will be available at the time of these measurements.

To check the robustness of our results under changes in the binning, we also performed $\Vcb$ fits for coarser grids: $7 \times 5$ bins for EIC while for LHeC and FCC-eh we divide each decade in $x$ in 6 bins and the $[0,1]$ $y$ interval in 16 bins.
The resulting $\Vcb$ sensitivity degrades by about a relative 2\% for the EIC and by less than one per mille for the higher energy colliders.

\begin{table}[t]
 \centering
 \small
 \begin{tabular}{lrrrrrr}
 \toprule
 \multicolumn{7}{c}{$\delta|V_{cb}|/|V_{cb}|\;[\%]$} \\
 \midrule
 Scenario & Stat. & CKM only & PDF only & Both & + top & + tags \\
 \midrule
 EIC $10\times275$, $200\,\mathrm{fb}^{-1}$ & 13.5 & 13.6 & 18.2 & 18.2 & & 21.9 \\
 EIC $18\times275$, $46\,\mathrm{fb}^{-1}$ & 18.9 & 19.0 & 23.5 & 23.5 & & 26.0 \\
 \midrule
 LHeC, $1\,\mathrm{ab}^{-1}$ & 0.269 & 0.312 & 0.439 & 0.478 & 0.488 & 0.789 \\
 FCC-eh, $1\,\mathrm{ab}^{-1}$ & 0.164 & 0.238 & 0.287 & 0.354 & 0.354 & 0.702 \\
 FCC-eh, $2\,\mathrm{ab}^{-1}$ & 0.116 & 0.179 & 0.225 & 0.304 & 0.304 & 0.668 \\
 \bottomrule
 \end{tabular}
 \caption{Nuisance dependence of the relative uncertainty on $\Vcb$ in percent, for negatively polarised electrons.}
 \label{tab:profiling}
\end{table}

The effect on the $\Vcb$ precision of the different nuisance groups is shown in Table~\ref{tab:profiling} for the polarised scenarios. 
We start with the statistical-only uncertainty, where all nuisance parameters are fixed at their reference values. 
Then we check the impact of floating the CKM elements and the PDF nuisance parameters, first each group independently and then together. 
We then add, only for the higher energy colliders, the 10\% top normalisation uncertainty and finally we include also the $C_0$ tagger calibration uncertainties of Table~\ref{tab:tagging}. 
For the EIC we see that statistical and systematic uncertainties are approximately of the same size.
Instead, at the LHeC and FCC-eh the statistical information is much greater, and calibration becomes central to the final sensitivity. Among these, current CKM uncertainties and the normalisation of the top background have a minor impact, compared to projected PDF uncertainties.
Tagger calibration uncertainties are instead the dominant source of the $\Vcb$ uncertainty.
This also explains why the improvement from electron polarisation is much smaller than what would be expected from the statistical scaling  $(1-P_e)^{-1/2}$ alone.

For the polarised LHeC baseline, taken as an example, we computed the post-fit nuisance constraints and correlations in the Gaussian approximation.
The largest correlations with $\mu_{cb}$ are with the efficiency calibrations $a_b$ ($-64\%$), $a_c$ ($-14\%$) and $a_{\rm light}$ ($-39\%$), as well as with $|V_{cs}|$ ($+27\%$). The strongest reduction in nuisance uncertainty occurs for the charm mistagging calibration $a_c$, whose posterior standard deviation is about $35\%$ of its prior value, followed by top normalisation ($66\%$) and light mistagging ($78\%$).

\subsection{Impact of the transverse-momentum threshold}
\label{sec:threshold}

\begin{figure}[t]
 \centering
 \includegraphics[width=0.95\textwidth]{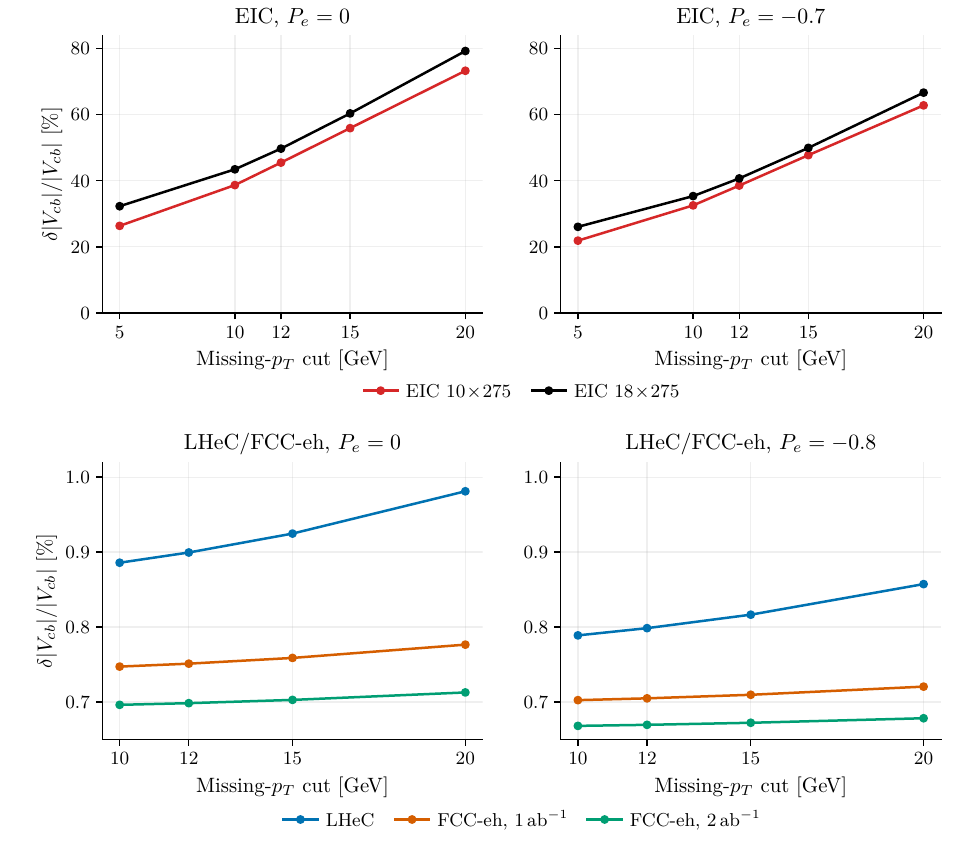}
 \caption{Dependence of the relative $\Vcb$ uncertainty on the $p_T$ threshold. All other cuts, projected PDFs, non-target CKM inputs and tag-calibration assumptions are those of the respective baseline.}
 \label{fig:threshold}
\end{figure}

A low transverse-momentum threshold is essential at the EIC. At Born level,
\begin{equation}
 \pt^2=xsy(1-y),\qquad x\geq\frac{4(\pt^{\rm cut})^2}{s}.
 \label{eq:threshold-bound}
\end{equation}
Increasing the threshold removes the low-$x$ region where the charm-initiated signal is most favourable relative to the valence-up background.
For example, a 20 GeV cut implies $x\gtrsim0.145$ at $10\times275$ GeV and $x\gtrsim0.081$ at $18\times275$ GeV, even before applying the remaining selections.
At the LHeC and FCC-eh the same cut probes much smaller $x$.

Figure~\ref{fig:threshold} shows how the $\Vcb$ sensitivity changes with the $p_T$ threshold for the different colliders and polarisation cases.
The binning, central tag probabilities and all other baseline nuisance assumptions remain fixed.

For the EIC $10\times275$ GeV stage at $P_e=-0.7$, increasing the threshold from 5 to 10 GeV worsens the $\Vcb$ uncertainty from $22\%$ to $33\%$, while at 20 GeV it reaches $63\%$.
The baseline EIC result therefore depends crucially on controlling the missing-momentum response and backgrounds in this low-threshold region.

The high-energy results are substantially less sensitive to the cut.
Raising it from 10 to 20 GeV changes the unpolarised LHeC error from $0.89\%$ to $0.98\%$, and the FCC-eh $2\,\mathrm{ab}^{-1}$ error from $0.70\%$ to $0.71\%$.
This weaker threshold dependence at the LHeC, as well as at FCC-eh, gives useful freedom to balance acceptance against experimental control without losing the main sensitivity.

\subsection{Impact of tagging efficiencies and calibration}
\label{sec:calibration}

\begin{figure}[t]
 \centering
 \includegraphics[width=0.95\textwidth]{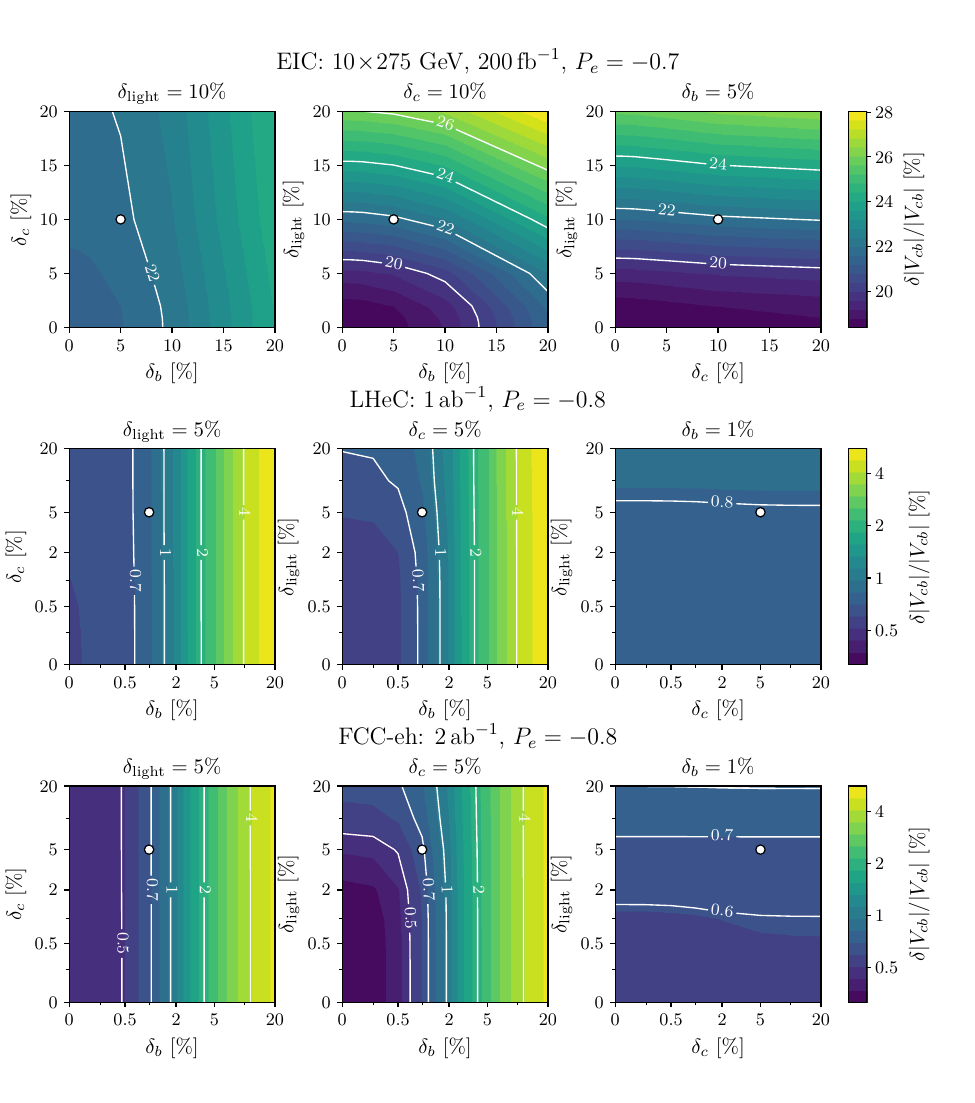}\vspace{-0.5cm}
 \caption{Dependence of the relative $\Vcb$ precision, in percent, on the tagging efficiency calibration uncertainties. The rows show results for EIC $10\times275$ GeV (top), LHeC with $1\,\mathrm{ab}^{-1}$ (central), and FCC-eh with $2\,\mathrm{ab}^{-1}$ (bottom). Only results for polarised electrons are shown.
In each plane, the omitted width is fixed to its baseline value $C_0$, and white circles mark the baseline.
EIC axes are linear while high-energy panels use coordinates $u=\log[1+\delta_f/(0.2\%)]$ with ticks labelled by the actual widths.}
 \label{fig:calibration}
\end{figure}

The signal rate is approximately proportional to $\epsilon_b\Vcb^2$. So, a relative bottom-efficiency uncertainty $\delta_b$ contributes roughly $\delta_b/2$ to the relative $\Vcb$ uncertainty.
The degeneracy is not exact because the denominator and other bottom-containing components also depend on the fitted parameters.
Nevertheless, a 1\% bottom-efficiency calibration already corresponds to a contribution of order 0.5\% in $\Vcb$.

To study how the final $\Vcb$ precision depends on these parameters, we perform a scan over the tagging efficiency calibration uncertainties $(\delta_b, \delta_c, \delta_{\rm light})$, keeping all central efficiencies unchanged as in Table~\ref{tab:tagging}.
We do this study for the EIC $10\times275$ GeV, LHeC, and FCC-eh $2\,\mathrm{ab}^{-1}$ cases, with polarised electrons. For each scenario, Figure~\ref{fig:calibration} shows three two-dimensional slices, each with the omitted calibration uncertainty fixed to its baseline value of $C_0$. 
The colours and white contours give profiled relative $\Vcb$ uncertainties. The white dots indicate our baseline $C_0$.

At the EIC, light-mistag calibration has the largest effect.
The larger anticharm yield does not make it the most limiting background, as its $(1-y)^2$ dependence helps distinguish it from the signal.
Instead, the light component remains more strongly correlated with the target after the other nuisances are profiled and so the corresponding calibration uncertainty plays the leading role.

On the other hand, at the higher-energy colliders bottom calibration is the leading tagging requirement.
As explained above, the reason is the very large statistics of the $b$ quark sample, which produces an approximate $\Vcb$ precision floor of order $\delta_b / 2$ in this setup.
Changing only the charm mistag uncertainty from 5\% to 20\% has very little effect while the light quark mistag calibration is only slightly more important, particularly at FCC-eh.

Next, we study how our results depend on the three central values of the tagging efficiencies.
The results are shown only for the LHeC and FCC-eh and only for negatively polarised electrons.
We consider two other tagging working points, in addition to the baseline UParT v2.
One comes from the ParticleNet B4 algorithm \cite{CMS:2025dsh} and the other is a hypothetical higher-performance tagging algorithm.
The corresponding efficiencies are collected in Table~\ref{tab:working-points}.
Each response is evaluated with two calibration uncertainties scenarios: the baseline $C_0$ and with all calibration widths halved, $C_0/2=(0.5,2.5,2.5)\%$.

\begin{table}[t]
 \centering
 \small
 \setlength{\tabcolsep}{4pt}
 \begin{tabular}{r|ccc}
 \toprule
  & $\epsilon_b$ & $\epsilon_c$ & $\epsilon_{\rm light}$  \\
 \midrule
 ParticleNet B4: & 0.39 & 0.001 & $10^{-4}$ \\
 UParT v2: & 0.60 & 0.0016 & $6\times10^{-5}$ \\
 Hypothetical: & 0.60 & 0.001 & $3\times10^{-5}$ \\
 \bottomrule
 \end{tabular}\\[5mm]
 
 \begin{tabular}{lrrrrrr}
 \toprule
 \multicolumn{7}{c}{$\delta|V_{cb}|/|V_{cb}|\;[\%]$} \\
 \midrule
 & \multicolumn{3}{c}{$C_0 = (1,5,5)\%$} & \multicolumn{3}{c}{$C_0/2 = (0.5,2.5,2.5)\%$} \\
 \cmidrule(lr){2-4}\cmidrule(lr){5-7}
 Collider, $\mathcal L$ & B4 & UParT & Hyp. & B4 & UParT & Hyp. \\
 \midrule
 LHeC, $1\,\mathrm{ab}^{-1}$ & 0.975 & 0.789 & 0.725 & 0.791 & 0.601 & 0.548 \\
 FCC-eh, $1\,\mathrm{ab}^{-1}$ & 0.934 & 0.702 & 0.639 & 0.651 & 0.479 & 0.441 \\
 FCC-eh, $2\,\mathrm{ab}^{-1}$ & 0.841 & 0.668 & 0.612 & 0.586 & 0.438 & 0.404 \\
 \bottomrule
 \end{tabular}
 \caption{Top: $b$-tagging efficiency scenarios.
 Bottom: relative $\Vcb$ uncertainty, in percent, for the three efficiency working points and two calibration assumptions, for negatively polarised electrons.}
 \label{tab:working-points}
\end{table}

The results in Table~\ref{tab:working-points} show that both rejection and calibration matter.
With $C_0$, ParticleNet B4 gives uncertainties $24$--$33\%$ larger than UParT.
Further improving the mistags to the hypothetical point reduces the UParT uncertainty by only about $8\%$.
On the other hand, halving the calibration uncertainties at fixed UParT gives a larger reduction of $24$--$34\%$.
Combining the improved rejection and halved calibration widths gives a $\Vcb$ uncertainty of $0.55\%$ ($0.40\%$) for the polarised LHeC (FCC-eh) case.
This comparison shows that improved calibration becomes much more important once tight rejection is available.

\subsection{Shape uncertainty in tagging efficiencies}
\label{sec:shape}

The global calibration factors studied above do not cover uncertainty in the kinematic dependence of the tag response.
Since the fit uses kinematic shapes to separate the signal from mistags, we introduce two additional nuisance parameters describing a residual $y$ dependence of the charm and light-quark mistag rates.
These variations preserve the total tagged yield of each flavour, separating them from the normalisation uncertainties already included.

Let us call $B_{fi}$ the tagged event yield of flavour $f= c,\mathrm{light}$ in bin $i$, and $y_i$ the midpoint of that bin in $y$.
For each $f$ we define a new rescaled basis for $y_i$, called $t_{fi}$, centred around the yield-weighted mean $\bar y_f$,
\begin{equation}
 \bar y_f=\frac{\sum_i B_{fi} y_i}{\sum_i B_{fi}} \, ,\qquad
 t_{fi}=\frac{y_i-\bar y_f}{\max_j|y_j-\bar y_f|} \, ,
 \label{eq:shape-basis}
\end{equation}
where the rescaled coordinates satisfy $|t_{fi}| \leq 1$ by definition.

For a standard Gaussian nuisance parameter $s_f$ and fixed relative width $\delta_{\rm shape}$, common to both flavours, the modified tagged event yield is
\begin{equation}
 B_{fi}(s_f) = B_{fi} \, e^{\delta_{\rm shape} \, t_{fi} \, s_f}
 \left(\frac{\sum_j B_{fj}}{\sum_j B_{fj}e^{\delta_{\rm shape} t_{fj} s_f}} \right)\,.
 \label{eq:shape}
\end{equation}
Note that the two nuisance parameters $s_c$ and $s_{\rm light}$ are independent, even though their widths are set to be equal for simplicity.
At the reference point the deformation is linear in $y$ to first order, while the exponential preserves positivity. The normalising factor in parentheses preserves the total yield for any value of $s_f$: $\sum_i B_{fi}(s_f) = \sum_j B_{fj}$.

\begin{figure}[t]
 \centering
 \includegraphics[width=0.95\textwidth]{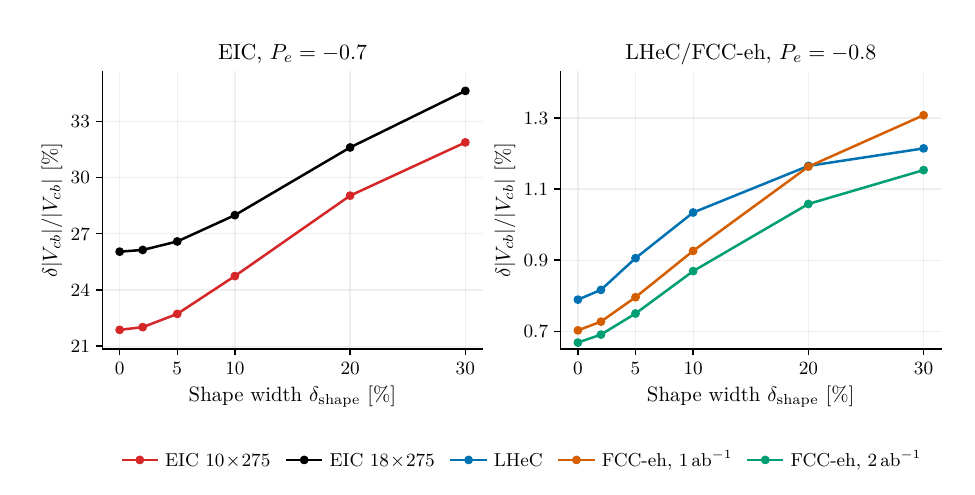}
 \caption{Dependence of the $\Vcb$ precision, in percent, on the mistag shape uncertainty width $\delta_{\rm shape}$. All other parameters remain as in the baseline fits.}
 \label{fig:shape}
\end{figure}

Figure~\ref{fig:shape} shows the dependence of the $\Vcb$ uncertainty  on the common width.
We see that LHeC and FCC-eh fits are more sensitive to this uncertainty than the EIC, as could be expected from the higher statistical precision.
At the EIC, a 2\% shape width increases the $\Vcb$ uncertainty by $0.2$--$0.7\%$ relative to the matched baseline while a 20\% shape width gives a $15$--$33\%$ increase.
For the LHeC, the uncertainty rises from $0.79\%$ at fixed shape to $0.82\%$, $0.91\%$ and $1.0\%$ at shape widths of 2\%, 5\% and 10\%, respectively.
Thus, percent-level control of relevant mistag shapes is necessary to preserve the sub-percent $\Vcb$ reach.

Although introduced as tagging-response uncertainties, these deformations also illustrate the sensitivity to analogous distortions of the tagged/inclusive ratios from other sources, including residual higher-order corrections and detector response. Their quantitative treatment, however, requires a dedicated study.

\subsection{Flavour tagging with SIDIS}
\label{sec:sidis}

\begin{figure}[t]
 \centering
 \includegraphics[width=0.6\textwidth]{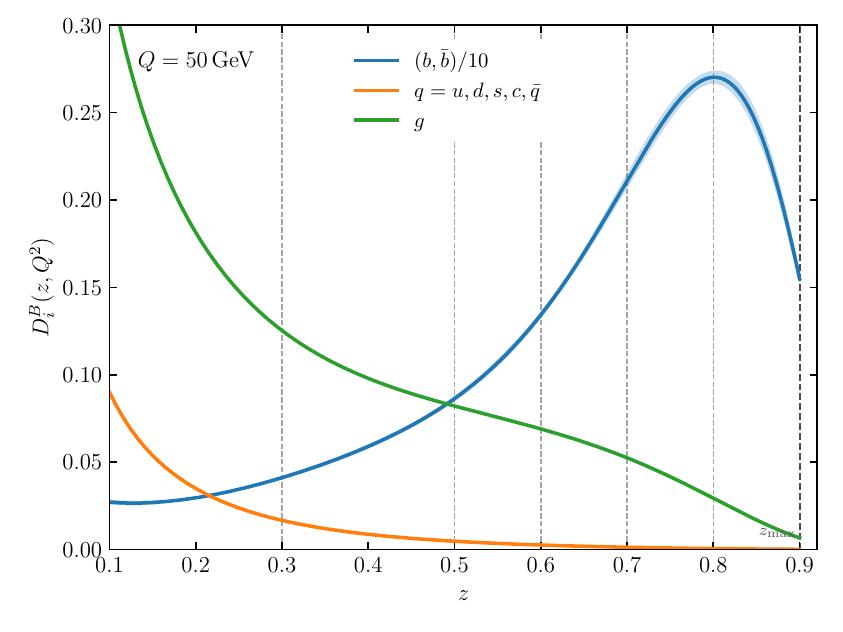}
 \vspace{-5mm}
 \caption{$B$ hadron FFs from Ref.~\cite{Czakon:2022pyz} evaluated at the scale $Q = 50$~GeV. Shading denotes the $1\sigma$ uncertainty. Vertical dashed lines identify our choices for the possible lower cuts $z_{\rm min}$ and the common upper cut $z_{\rm max} = 0.9$.}
 \label{fig:BFFshape}
\end{figure}

Semi-inclusive DIS is a process that requires the presence of a flavourful hadron in the final state, which carries a significant fraction of the hadronic energy. Such an energetic flavoured hadron is correlated with the flavour of the final-state parton in the hard DIS process, therefore SIDIS measurements offer a way to approach flavour tagging in DIS.

The correlation between a given flavourful hadron and the final-state quark is described via fragmentation functions, which can be seen as the final-state analogue of PDFs. We implement the NNLO $B$ hadron FFs from Ref.~\cite{Czakon:2022pyz}. 
It has one common FF for $q=u,d,s,c,\bar u,\bar d,\bar s,\bar c$ and equal $b$ and $\bar b$ FFs.
FFs correlated uncertainties in \cite{Czakon:2022pyz} are described as a set of 100 MonteCarlo replicas, which we compress to a set with a central member and 40 symmetric Hessian error directions via the {\tt mc2Hessian} method, analogously to what was done for the PDF sets.

The SIDIS variable $z$ indicates the fraction of the hadronic energy carried by the $B$ hadron.
In terms of the DIS kinematic variables introduced in Section~\ref{sec:processes} and the $B$ momentum $p_B$, it is defined as
$z=P\cdot p_B/(P\cdot q)$.
At the Born level, the FF can be interpreted as the probability to produce a $B$ hadron from a given final-state parton $i$ with a fraction $z$ of the initial partonic momentum.
The FFs, $D_i^B(z,Q^2)$, are represented in Figure~\ref{fig:BFFshape} for a reference scale $Q = 50$~GeV, with the corresponding 1$\sigma$ uncertainty bands.

We restrict the $z$ variable in a window $z \in [z_{\rm min}, z_{\rm max}]$, with no finer binning. The lower limit $z_{\rm min}$ is chosen a posteriori to maximise the sensitivity on $\Vcb$: we set $z_{\rm min} = 0.7$ at EIC and $z_{\rm min} = 0.5$ at LHeC and FCC-eh.
We define:
\begin{equation}
 \mathcal D_i^B(Q^2;z_{\min})=
 \int_{z_{\min}}^{0.9} dz\,D_i^B(z,Q^2) \, .
 \label{eq:ff-window}
\end{equation}
These integrals, at each $Q^2$, can be interpreted as the tagging efficiency and mistag rates. 
Here we report the FF integrals for two values of $z_{\min}$:
\begin{center}
 \centering
 \small
 \begin{tabular}{cccc}
 \toprule
 \multicolumn{4}{c}{$\mathcal D_i^B(Q^2;z_{\min})$, \quad $Q = 50$~GeV} \\
 \midrule
$z_{\min}$ & $b,\bar b$ & $q = u, d, s, c; \, \bar q$ & $g$\\
\midrule
0.5 & $0.759\pm 0.004$ & $(6.46\pm 0.06)\times10^{-4}$ & $(1.95\pm 0.01)\times10^{-2}$ \\
0.7 & $0.478\pm 0.005$ & $(1.11\pm 0.02)\times10^{-4}$ & $(5.84\pm 0.09)\times10^{-3}$ \\
\bottomrule
\end{tabular}
\end{center}
The gluon FF is not used in our Born level analysis. However, we note that it is much larger than those of the other partons, suggesting that it could give a significant background in a NLO analysis. While this goes beyond the scope of our work, tagging also the charge of the $B$ hadron could reduce its impact since the gluon contribution is charge-symmetric while our signal is not.

\begin{table}[t]
 \centering
 \small
 \begin{tabular}{l|r|r|r}
 \toprule
 \multicolumn{4}{c}{$\delta|V_{cb}|/|V_{cb}|\;[\%]$ - with $B$ FFs} \\
 \midrule
 Scenario & Perfect $B$-id & $(\epsilon_B^0, \delta_B)$ & Baseline\\
 \midrule
 EIC $10\times275$, $200\,\mathrm{fb}^{-1}$ & 8.3 & 14 & 22\\
 EIC $18\times275$, $46\,\mathrm{fb}^{-1}$ & 9.7 & 17 & 26 \\
 LHeC, $1\,\mathrm{ab}^{-1}$ & 0.67 & 0.97 & 0.79\\
 FCC-eh, $1\,\mathrm{ab}^{-1}$ & 0.60 & 0.90 & 0.70 \\
 FCC-eh, $2\,\mathrm{ab}^{-1}$ & 0.52 & 0.76 & 0.67 \\
 \bottomrule
 \end{tabular}
 \caption{Relative $\Vcb$ uncertainty in percent [\%] using $B$ hadron FFs instead of $b$-jet tagging, for negatively polarised electrons. The second columns assumes perfect $B$ identification, the third includes imperfect reconstruction efficiency and calibration (see text for details), and the last reports our baseline results with $b$-jet tagging.}
 \label{tab:sidis-vcb-results}
\end{table}

The SIDIS cross section is reported in Eq.~\eqref{eq:ff-xsec}. We integrate it in the same $(x,y)$ bins used in the main analyses and with those we construct the ratios to the inclusive DIS cross section, which are then used in the fit. 
For the top background we use the same selection as the main analysis. In this process, the FF is assigned to the jet's highest heavy flavour categoru ($b$ before $c$ before light), at $Q^2_{\rm vis} = s x_{\rm vis} y_{\rm vis}$.
All other fit parameters, except for those of the jet tagging, are the same as in the baseline fit. 

The resulting $\Vcb$ precision at each scenario and for negatively polarised electrons, after profiling over all the nuisance parameters (FFs ones included) and assuming a perfect $B$ hadron identification, is reported in the second column of Table~\ref{tab:sidis-vcb-results}. Comparing with the baseline result, reported for convenience in the last column, we see that the sensitivity improves across all scenarios. 
The largest improvement is for the two EIC scenarios, where the uncertainty is reduced by a factor of about 2.7, going below the 10\% level.

In reality, experimental identification of $B$ hadrons is not perfect.
We parametrise the corresponding reconstruction efficiency and calibration uncertainty similarly to what we did with the jet tagging efficiencies. The FFs integrals $\mathcal D_f^B(Q^2)$ are multiplied by a common efficiency factor $\epsilon_B(a_B) = \epsilon_B^0 e^{a_B \delta_B}$, see also Eq.~\eqref{eq:tag-response}.
Here we introduced the central $B$ reconstruction efficiency $\epsilon_B^0$, the relative calibration uncertainty $\delta_B$, and a new Normal nuisance parameter $a_B$.
For simplicity, and in absence of a dedicated experimental study on $B$ hadron identification, at each collider we set these parameters equal to the $b$-jet tagging efficiency and calibration uncertainties in the baseline scenario of Table~\ref{tab:tagging}: 
$(\epsilon_B^0, \delta_B)$ equal to $(0.35, 5\%)$ at EIC and $(0.60, 1\%)$ at LHeC/FCC-eh.
The $\Vcb$ precision obtained with these id efficiencies are reported in the third column of Table~\ref{tab:sidis-vcb-results}.
As expected, we see a reduction in precision compared to the perfect $B$-id case. However, while for the LHeC and FCC-eh scenarios the precision becomes worse than the one in the baseline, at the two polarised EIC scenarios it remains better, reaching 14\% at the 10$\times$275 stage with 200 fb$^{-1}$.

\subsection{Impact of the NC DIS background}
\label{sec:nc}

NC events can imitate a CC candidate if the scattered electron is poorly reconstructed, for instance if it hits close to uninstrumented detector regions. It should be noted that failure of identification alone, with the electron energy retained in the calorimeter, would not generate the required missing momentum.
Events in which the DIS scattering  directly involves a $b$ quark are particularly relevant to our $b$-tagged selection.
Quantifying this background requires modelling the full event selection and detector response, which is beyond the scope of our study.

We can derive an estimate from the HERA analysis of Ref.~\cite{H1:2012qti}. The NC cross section is one to two orders of magnitude larger than the CC one (Tables 45 and 47 of \cite{H1:2012qti}). After the selection, the NC background was found to be negligible (Section 5.4 of \cite{H1:2012qti}).
Assuming this means that it is at most $\sim 10^{-2}$ of the signal, we obtain that the probability of losing an electron is $p_{\rm loss} \sim 10^{-4}$.
A ZEUS analysis at HERA found negligible NC contamination after its CC-charm selection~\cite{ZEUS:2019oro}.
The LHeC detector proposal envisages electron vetoes and small-angle electron tagging to suppress NC and photoproduction backgrounds \cite{LHeC:2020van}.

We assess the size and possible impact of the NC background on the $\Vcb$ precision with a Born evaluation of the NC DIS process in the five-flavour scheme.
The scattered electron is assumed to be undetected with a uniform probability $p_{\rm loss}$, followed by ideal reconstruction of the remaining visible state.
The flavour-inclusive NC background contributes to the CC inclusive count, while we use the same $b$-tagging efficiencies as the baseline analysis to determine its contribution to the CC $b$-tagged event count.
We also include a 10\% relative uncertainty on $p_{\rm loss}$.

At the EIC, including this background gives a relative increase in the projected $\Vcb$ uncertainty of $1-2\%$ for $p_{\rm loss}=10^{-5}$ and $8-14\%$ for a $10^{-4}$ probability, compared to the baseline.
For the LHeC and FCC-eh scenarios, the relative increase remains around or below $1\%$ over $10^{-6}\leq p_{\rm loss}\leq10^{-3}$.
These results remain very similar upon changing the relative uncertainty on $p_{\rm loss}$ even up to 30\%.

In conclusion, within the assumed electron-loss model the small impact on the expected sensitivity supports neglecting this contribution in our baseline analyses. 
A precise experimental study on $\Vcb$ should however evaluate this background to avoid biasing the measurement itself.

\section{Conclusions}
\label{sec:conclusions}

The CKM matrix element $\Vcb$ is a crucial input for SM tests and indirect searches for new physics from rare flavour-changing processes.
Its current determinations from semileptonic $B$ decays have uncertainties above the 1\% level, dominated by theoretical and experimental systematic contributions.
Bottom-tagged charged-current DIS at electron-proton colliders offers an independent way to determine $\Vcb$ at high energy.
We have studied the prospects for such a measurement at the EIC, LHeC and FCC-eh using binwise tagged/inclusive ratios, with correlated PDF, CKM, and flavour-tag uncertainties.

At the EIC the expected relative precision could reach about 20\% with negatively polarised electrons and 200 fb$^{-1}$ of integrated luminosity at the $10\times 275$ GeV stage.
Although this precision would not match that of current determinations, a dedicated experimental analysis would provide an independent probe of $\Vcb$ in a different environment and help guide a future measurement at the LHeC.

Our main result is the LHeC sensitivity: with $1\,\mathrm{ab}^{-1}$ it could reach an expected uncertainty of $0.9\%$ for unpolarised electrons and $0.8\%$ at $P_e=-0.8$.
This would be competitive with current and future expected precision from semileptonic $B$-decay determinations and would probe the coupling with substantially different systematics.
If realised as proposed, the LHeC bridge programme could deliver this measurement before the FCC-ee $W$-decay programme.
FCC-eh provides more events and reaches the $0.7\%$ precision level, under the same tagging and calibration assumptions.

These results are supported by a different analysis done using the SIDIS process and $B$ hadron fragmentation functions instead of jet flavour tagging. Using this approach and assuming finite $B$ identification efficiency and calibration, we find similar results and even a possible improvement up to 14\% $\Vcb$ precision at the EIC.

Our analysis identifies concrete experimental requirements to realise this sensitivity level.
On the one hand, a bottom-tagging efficiency calibrated to about $1\%$ relative precision is important to reach the LHeC expectation. Reducing calibration widths by a factor of two improves the polarised uncertainty to $0.6\%$.
Control of the relevant mistag shapes at the few-percent level is also necessary.
On the other hand, raising the high-energy transverse momentum threshold to 20 GeV or widening the additional top normalisation uncertainty to 30\% has a modest impact on the final precision.
These comparisons make the LHeC case sufficiently promising to warrant a dedicated experimental study.

Our projected $\Vcb$ uncertainties should be seen as sensitivity targets.
A precise prediction must determine residual QCD and electroweak corrections to the tagged/inclusive observable, including the exclusive jet and balance requirements, consistently match heavy-flavour production channels, and evaluate radiative bottom CC production and photoproduction.
An experimental analysis must validate the low-$\pt$ and forward tag response, the reconstructed kinematic information, electron-loss rejection, and the treatment of physical top decays.

\section*{Acknowledgements}

We thank Juan Rojo for providing the projected EIC PDF sets, including the 50-direction Hessian representation.
AI tools were used in all stages of this work.

\appendix

\section{Kinematics and charged-current DIS cross sections}
\label{app:cross-sections}

\subsection{Two-body collider kinematics}

For the process $e^-(k)p(P)\to\nu_e(k')+X$, the main DIS variables are defined as in Eq.~\eqref{eq:dis-variables}.
The incoming massless quark carries momentum $p=\xi P$.
Setting the outgoing bottom quark on shell gives
\begin{equation}
 (\xi P+q)^2=m_b^2,\qquad \xi=x\left(1+\frac{m_b^2}{Q^2}\right),\qquad 0<\xi\leq1.
 \label{eq:xi}
\end{equation}
The measured Bjorken variable is $x$, while the PDF is sampled at $\xi$.

We take the proton along $+z$ and the electron along $-z$ and neglect their masses in the hard scattering. 
Writing $p^\pm=p^0\pm p^z$, the bottom momentum satisfies
\begin{align}
 p_b^+&=2E_p(\xi-xy) \, , & p_b^-&=2E_e y \, , \nonumber\\
 \pt^2(b)&=p_b^+p_b^- -m_b^2=Q^2(1-y) \, ,&
 p_{z,b}&=E_p(\xi-xy)-E_e y \, ,\\
 \eta_b&=\operatorname{arsinh}\!\left(\frac{p_{z,b}}{\pt(b)}\right) \, ,&
 \pt^{\rm miss}&=\pt(\nu_e)=\pt(b) \, . \nonumber
 \label{eq:massive-kinematics}
\end{align}
The hadronic-mass cut uses
\begin{equation}
 W^2=M_p^2+Q^2\frac{1-x}{x} \, ,
 \label{eq:hadronic-mass}
\end{equation}
retaining the target proton mass only in this variable.

\subsection{DIS cross sections}

The differential Born-level cross section for bottom-quark production is
\begin{equation}
 \frac{\dif^2\sigma_b^{\rm CC}}{\dif x\,\dif Q^2}
 =K(P_e,Q^2)\left[1+(1-y)\frac{m_b^2}{Q^2}\right]
 \left(\Vcb^2 f_c(\xi,Q^2)+\Vub^2 f_u(\xi,Q^2)\right)\Theta(1-\xi) \, ,
 \label{eq:bottom-sum}
\end{equation}
where we defined
\begin{equation}
 K(P_e,Q^2)=(1-P_e)\frac{G_F^2}{2\pi}
 \left(\frac{M_W^2}{M_W^2+Q^2}\right)^2.
 \label{eq:kernel}
\end{equation}
The square bracket in Eq.~\eqref{eq:bottom-sum} contains the massive flux and phase-space factor together with the change of variables from $\xi$ to $x$.
The massless non-top background DIS cross sections are, separating anticharm and light quarks:
\begin{align}
 \frac{\dif^2\sigma_{\bar c}^{\rm CC}}{\dif x\,\dif Q^2}
 &=K(P_e,Q^2)(1-y)^2\left(|V_{cd}|^2f_{\bar d}+|V_{cs}|^2f_{\bar s}+\Vcb^2f_{\bar b}\right),
 \label{eq:anticharm}\\
 \frac{\dif^2\sigma_{\rm light}^{\rm CC}}{\dif x\,\dif Q^2}
 &=K(P_e,Q^2)\left((|V_{ud}|^2+|V_{us}|^2)f_u+(|V_{cd}|^2+|V_{cs}|^2)f_c \right.\nonumber\\
 &\hspace{24mm} \left.+(1-y)^2\left[|V_{ud}|^2f_{\bar d}+|V_{us}|^2f_{\bar s}+\Vub^2f_{\bar b}\right]\right).
 \label{eq:light}
\end{align}
All PDFs in these two expressions are evaluated at $(x,Q^2)$.
We integrate the differential cross sections over the $(x,y)$ bins using
\begin{equation}
 \frac{\dif^2\sigma}{\dif x\,\dif y}=xs\frac{\dif^2\sigma}{\dif x\,\dif Q^2} \, .
 \label{eq:jacobian}
\end{equation}
LHAPDF supplies $\xi f_q(\xi,Q^2)$ for the massive signal and $xf_q(x,Q^2)$ for the massless backgrounds~\cite{Buckley:2014ana}, so we divide by the corresponding momentum fraction.

In the massless limit the differential SIDIS cross section is
\begin{align}
 \frac{\dif^2\sigma_{B-{\rm SIDIS}}^{\rm CC}}{\dif x\, \dif Q^2}
 &=K(P_e,Q^2)\left(
 \sum_{i=u,c}\sum_{j=d,s,b}|V_{ij}|^2 f_i(x,Q^2)
                  \mathcal D_j^B(Q^2)
 \nonumber \right.\\[-2pt]
 &\hspace{28mm} \left. +(1-y)^2
 \sum_{j=d,s,b}\sum_{i=u,c}|V_{ij}|^2 f_{\bar j}(x,Q^2)
                  \mathcal D_{\bar i}^B(Q^2) \right) \, .
 \label{eq:ff-xsec}
\end{align}

\section{Statistics}
\label{app:statistics}

For simplicity of notation, let $k_i$ be the observed tagged count and $n_i$ the observed inclusive count in bin $i$. The observed untagged count is $u_i = n_i - k_i$.
The likelihood for each bin is obtained by multiplying independent Poisson likelihoods of tagged and untagged events,
\begin{equation}
    \mathcal L_i = {\rm Pois}(k_i | N_i^\text{incl} R_i) \, {\rm Pois}(u_i | N_i^\text{incl}(1-R_i))~.
\end{equation}
This can be rearranged into a Poisson likelihood for $n_i$ and a binomial likelihood for the tagged fraction $R_i = N_i^{b\text{-tag}} / N_i^{\text{incl}}$,
\begin{equation}
    \mathcal L_i = {\rm Pois}(n_i | N_i^\text{incl}) \,  
        \binom{n_i}{k_i} R_i^{k_i} (1 - R_i)^{n_i - k_i}~.
\end{equation}
Since our observables are only the ratios, we drop the Poisson distribution for $n_i$.
In likelihood ratios constants cancel, so we can also drop the binomial coefficient. We thus arrive at our global likelihood
\begin{equation}
 \mathcal L(\mu_{cb},\bm\theta) \propto \prod_i \, R_i^{k_i}(1-R_i)^{n_i-k_i} \,
 \mathcal L_{\rm aux}(\bm\theta) \, .
 \label{eq:likelihood}
\end{equation}
The auxiliary likelihood is
\begin{equation}
 -2 \log \mathcal L_{\rm aux}(\bm\theta) = \bm\theta^{\mathsf T}\Sigma_{\rm aux}^{-1}\bm\theta = \sum_I \nu_I^2 + \sum_\alpha (V_\alpha - V_\alpha^{\rm ref})^2 / (\sigma_\alpha^{\rm PDG})^2 \,.
 \label{eq:nuisance}
\end{equation}
We evaluate the expected likelihood in Eq.~\eqref{eq:likelihood} with Asimov counts: $n_i = N_{i,0}^\text{incl}$ and $k_i = N_{i,0}^{b\text{-tag}} = N_{i,0}^\text{incl} R_{i,0}$~\cite{Cowan:2010js}.
The final expression for the test statistic is
\begin{equation}
 \Delta \chi^2 \equiv -2 \log \left(\frac{\mathcal{L}}{\mathcal{L}_0}\right) = 2 \sum_i N_{i,0}^\text{incl} \left[R_{i,0}\log\frac{R_{i,0}}{R_i}+(1-R_{i,0})\log\frac{1 - R_{i,0}}{1-R_i}\right]
 + \bm\theta^{\mathsf T} \Sigma_{\rm aux}^{-1} \bm\theta \, ,
\end{equation}
where $\mathcal{L}_0 \equiv \mathcal{L}(1, \bm 0)$.
To derive the profiled local Fisher uncertainty for $\delta\Vcb / \Vcb$ we compute the Fisher information matrix at the reference point:
\begin{equation}
 I_{ab} = \left. \frac{1}{2} \frac{\partial^2 \Delta \chi^2}{\partial \phi_a \partial \phi_b} \right|_{\rm ref} =
 \left. \sum_i\frac{n_i}{R_i(1-R_i)}\frac{\partial R_i}{\partial \phi_a}\frac{\partial R_i}{\partial \phi_b} \right|_{\rm ref}
 + I^{\rm aux}_{ab} \, ,
 \label{eq:fisher}
\end{equation}
where $\bm \phi = (\mu_{cb}, \bm \theta)$.
The profiled local Fisher relative uncertainty on $\Vcb$ is then
\begin{equation}
 \frac{\delta\Vcb}{\Vcb}\approx\frac{1}{2}\sqrt{(I^{-1})_{\mu_{cb}\mu_{cb}}}~.
 \label{eq:VcbUncertainty}
\end{equation}


\bibliographystyle{JHEP}
\bibliography{references}

\end{document}